\documentclass{SCIS2025}

\usepackage{tabularx}
\usepackage{boldline} 
\usepackage{multirow}
\usepackage{array}
\usepackage{threeparttable}
\usepackage{booktabs}
\usepackage{makecell}
\usepackage{multirow}
\usepackage[inkscapelatex=false]{svg}
\usepackage[numbers,sort&compress]{natbib}
\usepackage{float}
\usepackage{longtable}
\usepackage{array} 
\usepackage{booktabs} 
\usepackage{colortbl}
\usepackage{setspace}
\usepackage{enumitem}
\usepackage{hyperref}
\usepackage{breakurl}

\begin{document}
\ArticleType{RESEARCH PAPER}
\Year{2025}
\Month{January}
\Vol{68}
\No{1}
\DOI{}
\ArtNo{}
\ReceiveDate{}
\ReviseDate{}
\AcceptDate{}
\OnlineDate{}
\AuthorMark{}
\AuthorCitation{}

\title{Security and Privacy Measurement on Chinese Consumer IoT Traffic based on Device Lifecycle}

\author[]{Chenghua~Jin}{}
\author[]{Yuxin~Song}{}
\author[]{Yan~Jia}{jiay@nankai.edu.cn} 
\author[]{Qingyin~Tan}{}
\author[]{\\Rui~Yang}{}
\author[]{Zheli~Liu}{}


\address[1]{Key Laboratory of Data and Intelligent System Security, Ministry of Education, China (DISSec), \\Tianjin Key Laboratory of Network and Data Security Technology (NDST), College of Cryptology and Cyber Science,\\Nankai University, Tianjin 300350, China. }

\abstract{In recent years, consumer Internet of Things (IoT) devices have become widely used in daily life. With the popularity of devices, related security and privacy risks arise at the same time as they collect user-related data and transmit it to various service providers. Although China accounts for a larger share of the consumer IoT industry, current analyses on consumer IoT device traffic primarily focus on regions such as Europe, the United States, and Australia. Research on China, however, is currently relatively rare. This study constructs the first large-scale dataset about consumer IoT device traffic in China. Specifically, we propose a fine-grained traffic collection guidance covering the entire lifecycle of consumer IoT devices, gathering traffic from 77 devices spanning 38 brands and 12 device categories. Based on this dataset, we analyze traffic destinations and encryption practices across different device types during the entire lifecycle and compare the findings with the results of other regions. Compared to other regions, our results show that consumer IoT devices in China rely more on domestic services and overall perform better in terms of encryption practices. However, there are still 23/40 devices improperly conducting certificate validation, and 2/77 devices use insecure certificate algorithms. To facilitate future research, we open-source our traffic collection guidance and make our dataset publicly available.

}

\keywords{Chinese Consumer IoT, Traffic Dataset, Network Traffic Measurement, Privacy, Security}

\maketitle


\section{Introduction}
\label{sec:intro}

The consumer Internet of Things (IoT) industry has rapidly developed in recent years, with various devices becoming widespread and offering numerous conveniences. According to the report released by China Internet Network Information Center~\cite{CNNIC} in 2025, as of November 2024, the number of IoT end-users had increased to 2.642 billion. However, the rapid spread of devices has also raised growing concerns regarding security and privacy risks~\cite{SCIS}. According to the threat report jointly released by Bitdefender and Netgear in 2024~\cite{BitdefenderNetgear}, IoT devices, such as TVs, hubs, and cameras, are subjected to over 2.5 million attack attempts on average every day. As consumer IoT devices usually have the ability to collect and transmit data, the data collected by these devices is often transmitted to the manufacturers of these devices and other service providers. At the same time, data transmission is facing strict legal regulations~\cite{SecurityLaw} in China. All data collected and transmitted by these devices is contained within the traffic they generate, therefore, it is necessary to measure consumer IoT traffic from security and privacy perspectives.

Researchers have conducted studies on consumer IoT device traffic in various regions, including Europe~\cite{7980220, bhandari2023distributed}, the United States~\cite{10.1145/3355369.3355577, osti_10314043, alrawi2019sok}, and Australia~\cite{8440758, koroniotis2019towards}, and have identified several security and privacy risks associated with these devices. However, according to IDC~\cite{IDC}, China’s consumer IoT industry is projected to reach \$300 billion by 2027. Despite this rapid growth, research on consumer IoT device traffic in China remains limited and lacks a comprehensive analysis. The research in China has not made their data publicly available~\cite{9155516}, and only contains the traffic of the camera. This research gap drives us to build a large-scale and comprehensive dataset of consumer IoT traffic in China and conduct a measurement of the dataset.

While constructing the dataset, we introduce methodological improvements compared to existing studies. Previous studies have primarily focused on the device interaction phase~\cite{10.1145/3355369.3355577,paracha2021iotls}. However, we observe differences in traffic between different operational phases of the device, and focusing only on the interaction phase, may lead to incomplete analysis results. Therefore, we propose a fine-grained traffic collection guidance that covers the entire lifecycle of the device to address this limitation. The entire lifecycle refers to all phases during normal device usage, including device setup, user interaction, idle, and deletion~\cite{Mao2019lifecycle}. Compared with prior works, the traffic we collected shows greater coverage and can support more fine-grained analysis. Furthermore, we open-source our traffic collection guidance and make our dataset publicly available to facilitate further research.

According to our traffic collection guidance, we collected the first part of device traffic between March 1 and May 30, 2024, covering 50 consumer IoT devices. To further expand the dataset, we conducted two additional rounds of traffic collection involving 27 more devices: the second from November 1 to November 29, 2024, and the third from June 7 to June 19, 2025. During our 4.5 months traffic collection experiment, we compiled a dataset of 108 hours of traffic from 77 consumer IoT devices in China, the total size of the dataset is 3 GB. Compared with the existing datasets, our dataset covers the full lifecycle traffic of the devices and contains the largest number of devices within a single region. Furthermore, For devices capable of firmware updates during the experiment, we also collected traffic after the update to compare the differences pre- and post-update. Based on the traffic collected, we analyzed the traffic destinations, encryption practices, and then compared our findings with those of the United States and the United Kingdom~\cite{10.1145/3355369.3355577,paracha2021iotls}. 

Our study reveals that, regarding traffic destinations, Chinese devices show a higher reliance on domestic services. The distribution of traffic destinations for Chinese devices is mainly directed to China (99.75\%) and relies on domestic organizations. In contrast, devices from the United Kingdom have a significantly higher proportion of overseas traffic, with a greater reliance on organizations from the United States. Regarding encryption practices, devices from different regions all show great encryption practices, although some issues remain. In particular, Chinese devices have less PII exposure in unencrypted traffic and a lower percentage of unencrypted traffic compared to devices in the United States and the United Kingdom. However, 23/40 devices still use improper certificate verification, and 2/77 use insecure certificate algorithms.

Our main contributions include:

\begin{itemize}[itemsep=2pt,topsep=0pt,parsep=0pt]
    \item \textbf{The first Chinese consumer IoT devices traffic dataset.} We constructed a dataset that includes 108 hours of traffic from 77 devices across 12 categories and 38 brands. We have publicly released the dataset\footnote{\url{https://github.com/NKUHack4FGroup/Lifecycle-Based-Traffic-Dataset}} to support future research within the community.
    \item \textbf{A lifecycle-based fine-grained traffic collection guidance.} We propose a fine-grained traffic collection guidance covering the entire lifecycle of the device in a controlled environment. We open-sourced the guidance and invite researchers to build a standard, comprehensive, and state-of-the-art consumer IoT dataset together.
    \item \textbf{Security and privacy measurement of Chinese consumer IoT device traffic.} 
    Based on the dataset, we developed scripts to analyze the traffic destinations and encryption practices, and compared the results with those from other regions, revealing the characteristics of Chinese consumer IoT devices and regional differences. We have made the source code available online.
\end{itemize}

\section{Background}
\label{sec:back}

Consumer IoT devices typically refer to IoT devices designed for personal or household use, such as cameras, speakers, and lights. These devices can be bound to and controlled by the companion apps, following similar operational phases (i.e., lifecycle). During these phases, users can control the devices through various methods. Meanwhile, consumer IoT devices communicate frequently with cloud servers during operation to support various device functions. This section introduces the concepts and definitions relevant to the lifecycle, control methods, and cloud servers mentioned above.

 \textbf{The lifecycle of consumer IoT devices.} Based on a prior work~\cite{Mao2019lifecycle}, the lifecycle of consumer IoT devices contains four phases: setup, interaction, idle, and deletion. When users first activate a consumer IoT device, the device will enter the \emph{setup phase}. During this phase, users need to complete operations such as network configuration and bind the device to the companion app. The \emph{interaction phase} refers to the phase in which users interact with the device, such as controlling light switches and checking camera status. After setting up the device, we refer to the phase in which the user no longer interacts with the device as the \emph{idle phase}. Finally, the \emph{deletion phase} refers to the phase when users unbind the device from the companion app, restoring it to its factory settings.

 \textbf{Control methods of consumer IoT devices.} Common control methods of consumer IoT devices include four types: Wide Area Network (WAN) control, Local Area Network (LAN) control, physical control, and multimodal control. \emph{WAN control} refers to scenarios where the controlling smartphone with the companion app and the consumer IoT device are not on the same local network. In this case, commands between the app and the device must be relayed by a cloud server. For example, when users are away from home during travel, they can monitor the indoor environment in real-time using a camera via a companion app. \emph{LAN control} occurs when the controlling smartphone and the consumer IoT device are on the same local network. In this case, communications between the device and the app can be established by using local protocols, such as Bluetooth. For example, when users are at home, they can use the companion app to adjust the brightness of smart lights. \emph{Physical control} involves direct interaction with the device by using the physical buttons. For example, users can press physical switch buttons on the device to reset it or power it off. Consumer IoT devices with various sensors support \emph{multimodal control} through different methods. Examples include users waking up speakers by using voice commands or interacting with devices via triggers from motion sensors.

 \textbf{Cloud servers of consumer IoT devices.} 
 When a consumer IoT device is in operation, the data it collects will be shared with various cloud servers. Following the classification method used in the study by Ren et al.~\cite{10.1145/3355369.3355577}, we categorize these servers into three groups: first party, support party, and third party. \emph{First party} refers to the manufacturer of the consumer IoT device or the developer of its companion app. \emph{Support party} refers to organizations that provide computational resources or network services for consumer IoT devices, such as cloud computing, cloud storage service providers, and service providers (e.g., CDN). \emph{Third party} refers to companies contacted by the device and not categorized as first or support parties. We adopt the methodology from Ren et al.~\cite{10.1145/3355369.3355577} to determine the server's party through a three-step verification process: (1) We first identify the organization of the IP addresses contacted by the device. If the IP’s organization matches the manufacturer or companion app of the consumer IoT device, we classify it as a first party server; (2) For unresolved cases, examine official Privacy Policies and Third-Party Data Sharing documents (prioritizing device-specific policies) to identify the support party server; (3) Categorize any undocumented or unaffiliated servers as the third party server.

\section{Traffic collection methodology and measurement scope}
\label{sec:Data Collection Methodology}

This section introduces the traffic collection methodology and measurement scope in our study. The traffic collection methodology includes the device selection criteria, experiment testbed configuration, and traffic collection guidance. The measurement focuses on two main dimensions: traffic destinations and encryption practices. We collected the device traffic between March 1 and June 1, 2024, November 1 to November 29, 2024, and June 7 to June 19, 2025. During the traffic collection experiment, a total of 108 hours of traffic were systematically collected from 77 Chinese consumer IoT devices. We also collected traffic from ten devices after firmware updates to study differences in device behavior before and after firmware updates, generating 26 hours of traffic data covering the pre- and post-firmware updates.

\subsection{Experiment testbed}

\begin{table}[h]
\Huge
\centering
\caption{Device list.}
\label{table:devicelist}
\resizebox{\textwidth}{!}{
\begin{tabular}{c|cccccccccl}
\toprule[4pt]
\textbf{Category}                                      & \multicolumn{2}{c}{\textbf{Cameras (29)}}                                                                                                                                                                                                                                                                                                                                                       & \textbf{Doorbells (3)}                                             & \textbf{Hubs (9)}                                                                                                        & \textbf{Humidifiers (2)}                                & \textbf{Lights (6)}                                                                                          & \textbf{Plugs (12)}                                                                                                                        & \textbf{Sensors (3)}                                            & \textbf{Speakers (9)}                                                                                                                                                                              & \textbf{Other devices (4)}                                                                                             \\ \hline
\begin{tabular}[c]{@{}c@{}}Devices\\ (77)\end{tabular} & \begin{tabular}[c]{@{}c@{}}XiaobaiY2 \\ XiaobaiY3\\ Xiaomi\\ Huawei haique\\ Xiaopai\\ Cubetoou\\ ZTE\\ Yingshi C6CN\\ 360 6C\\ Jooan\\ Konka\\ Pisen\\ Lenovo\\ Yibang\end{tabular} & \begin{tabular}[c]{@{}c@{}}Tenda\\ Mercury\\ Aqara\\ Xiaovv\\ Huawei xiaotun\\ Xiaotundangjia\\ TP-Link\\ IMOU\\ Greatwall\\ Xiaomi 3\\ JPLAYER\\ Haier\\ Sonoff\\ Xiaomao\\ Xiaobai Smart\end{tabular} & \begin{tabular}[c]{@{}c@{}}IMOU \\ Chuangmi\\ TP-LINK\end{tabular} & \begin{tabular}[c]{@{}c@{}}Aqara\\ Yingshi\\ Mindor\\ Tuya\\ Xiaomi\\ Jingxun\\ Meian\\ Sonoff\\ BroadLink\end{tabular} & \begin{tabular}[c]{@{}c@{}}Smartmi\\ Mijia\end{tabular} & \begin{tabular}[c]{@{}c@{}}Huawei dalen\\ Wiz\\ Philips\\ Mijia\\ Aqara\\ Midea\end{tabular} & \begin{tabular}[c]{@{}c@{}}Hongyar\\ Mijia2\\ Mijia3\\ Gosund\\ Yingshi\\ Mindor\\ Tuya\\ Xiaodu\\ Xiangrikui\\ Bull\\ Aqara\end{tabular} & \begin{tabular}[c]{@{}c@{}}Tuya\\ Yingshi\\ Xiaodu\end{tabular} & \begin{tabular}[c]{@{}c@{}}Midea Speaker\\ Huawei Speaker\\ Xiaomi Speaker\\ Xiaomi Speaker Plus\\ Tmall Speaker\\ Xiaodu Clock\\ Jingyuzuo Clock\\ Xiaodu Speaker\\ XiaomiTV Speaker\end{tabular} & \multicolumn{1}{c}{\begin{tabular}[c]{@{}c@{}}Dogness Feeder\\ Deebot Cleaner\\ Midea Mirror\\ Huawei TV\end{tabular}} \\ \toprule[4pt]
\end{tabular}}
\end{table}
We divide the experiment environment into LAN and WAN to capture device traffic under different network scenarios. To ensure diverse and comprehensive traffic collection, we selected a total of 77 consumer IoT devices across 12 categories (cameras, plugs, speakers, hubs, humidifiers, lights, doorbells, sensors, mirror, TV, Cleaner, and pet feeder). To ensure the representativeness of adopted consumer IoT devices, we prioritized brand recognition, user reviews, and ratings on major Chinese e-commerce platforms (e.g., Taobao, JD.com) during the selection process. The complete list of devices and corresponding brand information is provided in Table~\ref{table:devicelist}.

\begin{figure}[h]
\centering
\includegraphics[width=0.5\linewidth]{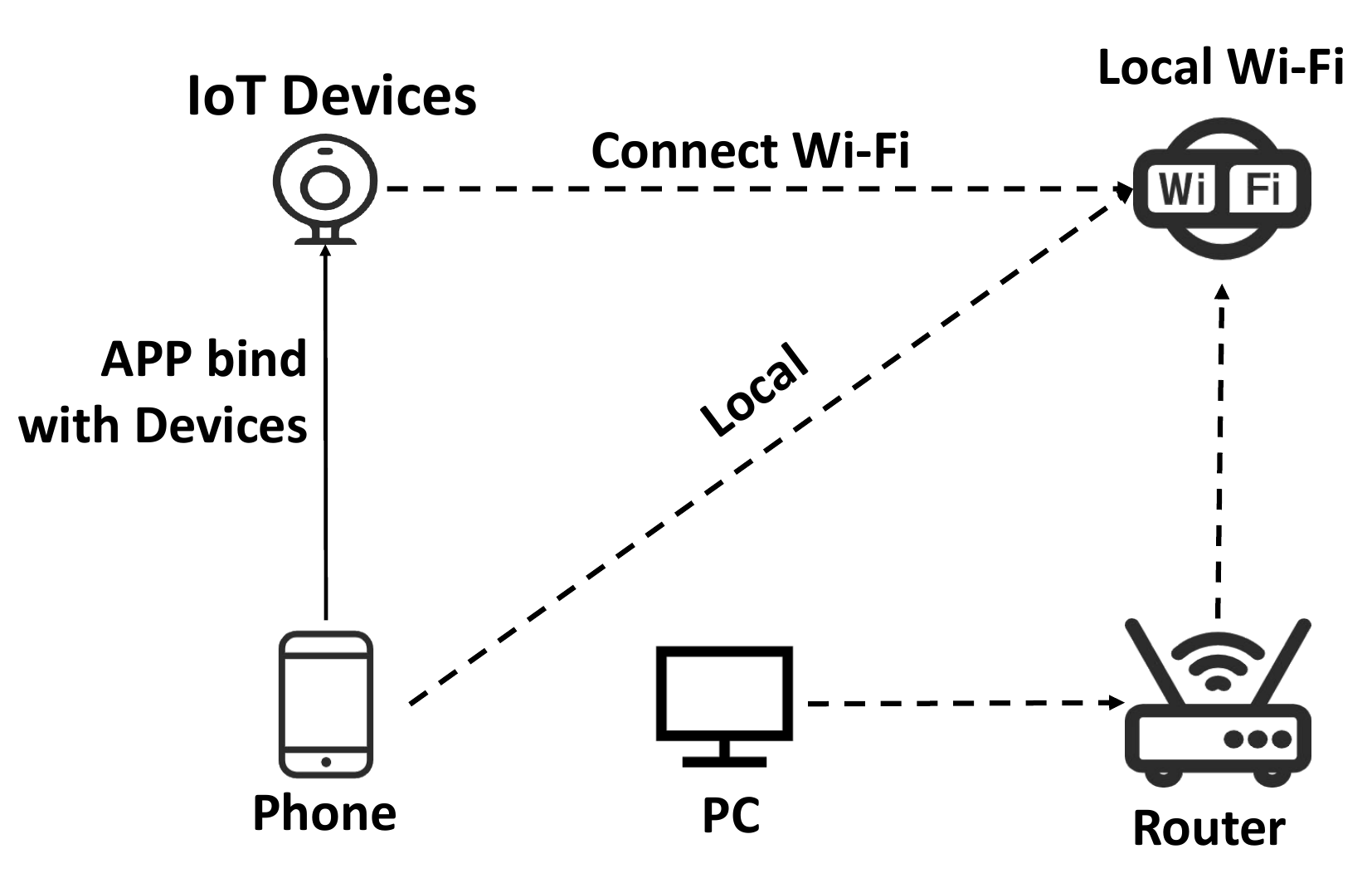}
\caption{The setting of the controlled experiment environment.}
\label{Fig:Experimental_topology}   
\end{figure}

To minimize external interference, experiments are conducted in a controlled environment. The setting of the controlled experiment environment is shown in Figure~\ref{Fig:Experimental_topology}. The router in the figure is a Xiaomi AX9000 running OpenWrt, with Tcpdump installed on both the WAN and LAN interfaces for traffic capture. An Android smartphone, with the latest versions of companion apps downloaded from the corresponding manufacturer's official websites, is used during the experiments. For LAN operations, both the smartphone and devices are connected to the local area network, while WAN operations require the smartphone to disconnect from the LAN. All devices are configured according to the standard setup process and bind to their respective companion apps. A Windows 10 PC is employed to configure the router and collect the traffic throughout all phases. The traffic collection process follows a semi-automated approach, and the experimenters adhere to the Traffic Collection Guidance~(section~\ref{3.2}). This guidance combines automated scripts with manual operations, ensuring flexibility in the experiment while simulating realistic scenarios.

\subsection{Traffic collection guidance}
\label{3.2}

Previous studies have mainly relied on manual methods for traffic collection~\cite{trimananda2020packet}. While effective, they often lack a standard operating procedure. To address these limitations, Ren et al.~\cite{10.1145/3355369.3355577} proposed a fully automated click-based method to reduce human intervention. However, some operations, like physical controls, remain challenging to automate entirely, and complete automation fails to replicate real-world user behavior accurately. To prevent the effect of personal operations and guarantee comprehensive traffic collection, we propose a semi-automated traffic collection guidance. Our traffic collection guidance follows a semi-automated design principle comprising four core files:

\begin{itemize}[itemsep=2pt,topsep=0pt,parsep=0pt]
    \item \textbf{Manual operation process files.} The file predefined the operations that the experimenters need to perform for each type of device.
    \item \textbf{Timestamp files.} Record the operations performed on the device and the corresponding times.
    \item \textbf{Interactive traffic collection script.} Start Tcpdump to collect traffic and automatically remind the experimenters to perform corresponding operations according to the manual operation process files. During the collection, the operations and the corresponding times will be recorded in timestamp files. After collecting the traffic, the raw traffic will be saved.
    \item \textbf{Traffic segmentation script.} Split the raw traffic according to Timestamp files.
\end{itemize}

Upon initiating traffic capture, the collection script starts Tcpdump and loads the manual operation process file. Experimenters are guided by operation prompts to perform predefined operations, such as device setup and firmware updates. The start and end times of each operation are automatically logged in the timestamp file. After traffic collection, the segmentation script processes the raw PCAP file using the timestamp file, automatically splitting it into multiple lifecycle phases for analysis. Additionally, there are 13\% of all devices that supported firmware updates during the experiment, including five cameras (Aqara camera, TP-Link camera, XiaobaiY2 camera, Xiaomi camera), four hubs (Aqara hub, Boardlink hub, Xiaomi hub, Yingshi hub), and one plug device (Mijia plug). For these devices, we collected traffic both before and after the update process. After the update, devices are re-bound to their companion apps, and traffic is recaptured using the same process. This semi-automated approach ensures comprehensive coverage of the device lifecycle, accurately reflecting real-world controls. Following the established environment and traffic collection guidance, we successfully collected 108 hours of traffic, including 13 hours of post-firmware-update traffic from 10 devices.

\subsection{Measurement scope}

The consumer IoT devices continuously collect large amounts of data about users and their surrounding environments during daily use. Much of this data may contain sensitive personal information, making it crucial to evaluate how these devices transmit data and to what destinations. Our study focuses on the security and privacy practices of these devices by analyzing both the traffic destinations and encryption~practices.

First, we analyzed the traffic destinations of the devices. Given the location of the devices, all the devices analyzed in this study are popular products in China, therefore, we focus on traffic directed to destinations outside of China. Through traffic analysis, we assess where devices send data, the proportion of domestic versus overseas traffic, and the services provided by these destinations. Additionally, we further analyze which parties the service providers belong to because the traffic that goes to non-first parties tends to have a higher risk than traffic that goes to first parties. Moreover, we identified several domains frequently accessed by most devices, which were linked to specific organizations. By analyzing these domains, we identified the organizations most frequently contacted by the devices. This analysis sheds light on the reliance of consumer IoT devices on different organizations.

Second, we conduct a detailed analysis of the encryption practices of device traffic. First, we measure the proportion of encrypted traffic. We then categorize the types of encryption protocols used in the encrypted traffic. Due to TLS 1.2 is widely used in device traffic, we conduct an in-depth analysis of the security of certificates using TLS 1.2, focusing on whether the algorithms used in the certificates meet current security standards and whether the device verifies the certificate correctly. For unencrypted traffic, we assess the potential risk of exposing personally identifiable information (PII) in plain text, with particular attention to the potential leakage of sensitive information.

\vspace{-1em}

\section{Destination analysis}
\label{sec:Destination Analysis}

For consumer IoT devices in China, traffic directed to overseas locations warrants attention. Similarly, traffic flows to third party servers generally present higher security and privacy risks compared to traffic flows to first party or support party servers. This section starts with an analysis of traffic destinations, providing a detailed assessment of security and privacy risks through two key aspects: (1) the geographical distribution of traffic destinations by country and (2) the analysis of contacted server parties.

\subsection{Distribution of traffic destinations}

This subsection analyzes the country-level distribution of device traffic destinations, focusing on the proportion of traffic to different countries in different device types and lifecycle phases. 

We first collected IP addresses interacting with devices and queried their geographical locations using the IP geolocation API ~\cite{IPapi}. Then, we calculate the proportion of traffic going to different countries. In the calculation, we observed that the traffic packet size of cameras, doorbells, and speakers is significantly larger than that of sensors and other devices. If only the proportion of bytes flowing to devices in different countries is counted, the final result will largely depend on the traffic destinations of devices such as cameras, and the influence of sensors and other devices will be negligible, resulting in a bias. To eliminate the above bias, we processed a new calculation method. 

Let \( X \) represent the set of type \( A \) device indices in the dataset. For each device \( i \) (\( i \in X \)), let \( M_i \) denote the traffic (in bytes) transmitted to the country \( T \), and let \( N_i \) denote the total traffic transmitted to all destinations. The total normalized traffic from type \( A \) devices to country \( T \) is then given by \(A_T = \sum_{i \in X} \frac{M_i}{N_i}\). Further, let \( Sum_A \) represent the total traffic of all type \( A \) devices. The proportion of traffic from type \( A \) devices directed to country \( T \) is computed as:  \(\frac{A_T}{Sum_A}\). Similarly, let \( Sum_T \) represent the total traffic from all devices to the country \( T \), and \( Sum_{all} \) denote the total traffic from all devices to all destinations. The overall proportion of traffic directed to the country \( T \) across all devices is then given by:  \(\frac{Sum_T}{Sum_{all}}\).

\begin{figure}[h]
\centering
\includegraphics[width=0.9\linewidth]{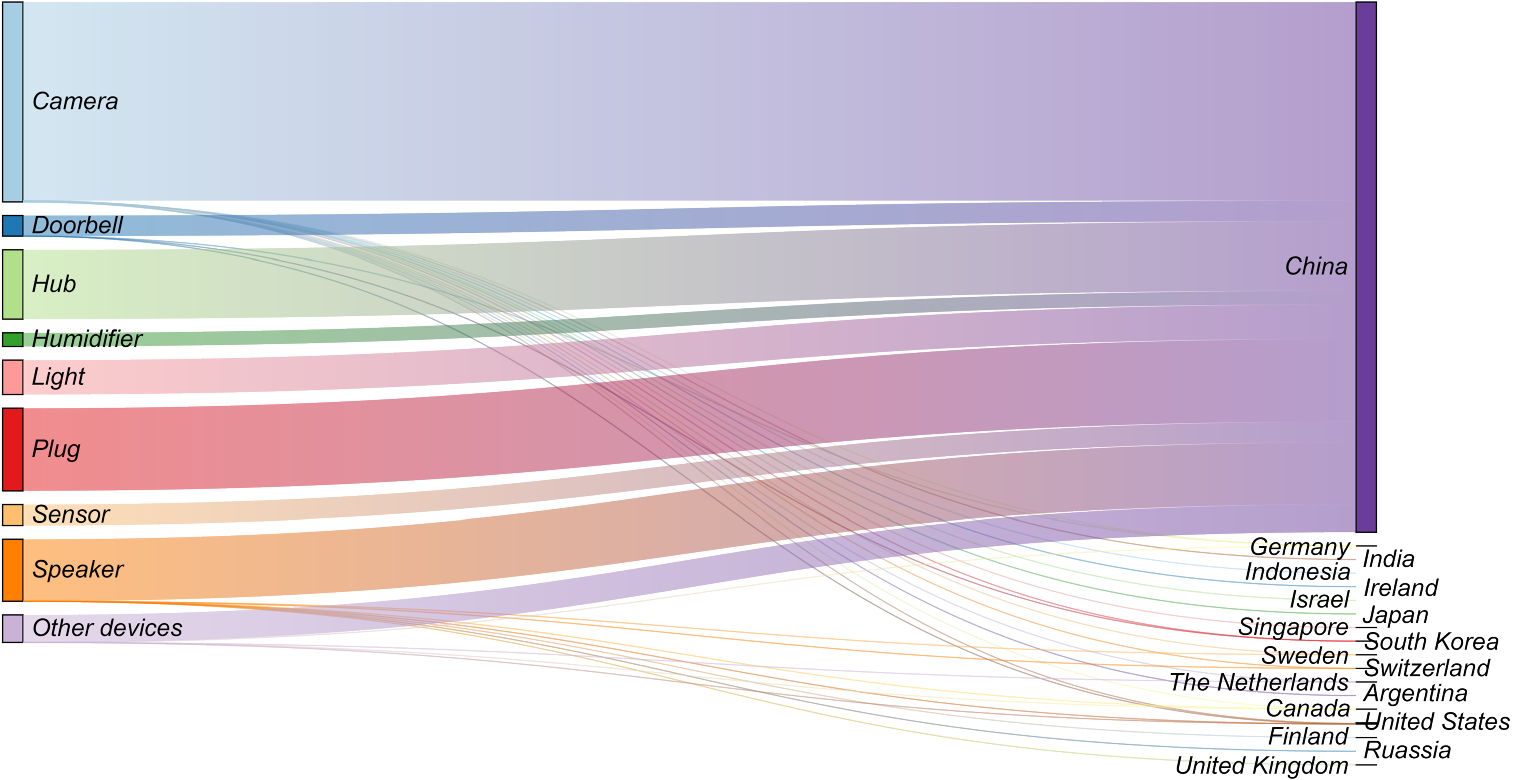}
\caption{Proportion of device traffic routed to different countries, grouped by device type and traffic destination country.}
\label{Fig:destination_sankey}
\end{figure}

We used Chiplot~\cite{Chiplot} to generate a Sankey diagram (Figure~\ref{Fig:destination_sankey}) to visualize the distribution of traffic destinations. In terms of destination, 99.75\% of all device traffic remained within China, with only 0.25\% directed overseas. From the perspective of device types, hubs, humidifiers, lights, sensors, and plugs did not show overseas traffic, likely due to their simple communication with cloud servers. A total of 32 devices showed overseas traffic, including cameras, speakers, doorbells, TV, mirror, and pet feeder. For each type of device, the cameras had the highest proportion of overseas traffic at 0.6\%, followed by speakers at 0.03\%, doorbells at 0.01\%, and the other devices at 0.02\%. 

The United States is the most common overseas destination, accounting for 0.24\% of total traffic. In particular, \url{dns.google.com} is the most frequently accessed overseas domain, communicated by 21 devices. Other domains involved in the United States traffic include \url{ntp.com}, \url{amazonaws.com}, and \url{iotcplatform.com}, supporting services such as time synchronization (e.g., \url{1.cn.pool.ntp.org}), remote access based on TUTK (e.g., \url{all-d-master-tutk.iotcplatform.com}, used by Xiaobai doorbells and Jooan cameras), and cloud computing (e.g., \url{p2ps2.coolkit.cn}, accessed by Sonoff cameras).
Traffic to other countries accounted for less than 0.01\% of the total and involved 18 domain names. Among them, 10 domains (e.g., \url{ntp.org}, \url{time.cloudflare.com}) are related to NTP services.  Other services included Google Cloud (e.g., \url{bc.googleusercontent.com}), IoT platform management (e.g., \url{all-master.iotcplatform.com}), IP address lookup (e.g., \url{api.ipify.org}), and AWS cloud hosting (e.g., \url{eu-central-1.compute.amazonaws.com}).

From the perspective of cross-border data transmission security, Chinese consumer IoT devices still rely on overseas infrastructure services, which may violate China's data transmission security regulations. Platforms behind services such as ``iotcplatform.com'' and ``amazonaws.com'' depend on complex third party components. This kind of dependency increases the possibility of unauthorized cross-border data flows. Users’ sensitive data may be sent to overseas servers without clear consent, raising the risk of data leaks and misuse. In addition, such cross-border dependencies may affect device availability. For example, network disruptions or policy changes could limit access to remote services, affecting the stability of device functions. Therefore, device manufacturers should consider adopting service localization strategies and reliable domestic infrastructure providers. 

It should be noted that 29 devices contacted NTP servers in 12 countries, and 8 devices contacted NTP servers in multiple countries. In contrast, domestic NTP services could offer more stable connections.
Therefore, we suggest device manufacturers prioritize the use of domestic NTP sources during product design. This approach reduces potential risks from cross-border access and elevates the overall security protection capabilities of the system.

\begin{figure}[h]
\centering
\includegraphics[width=0.75\columnwidth]{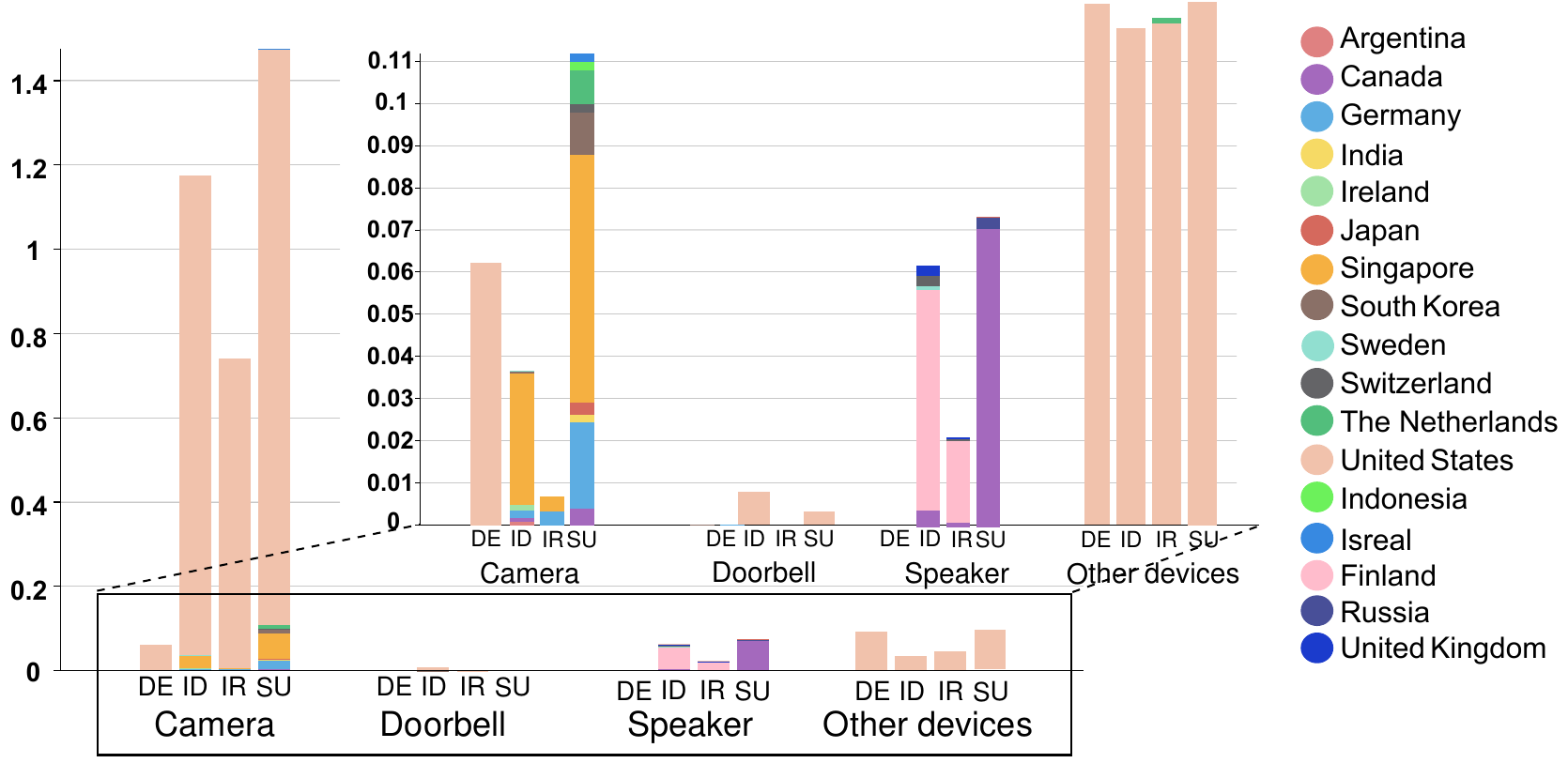}
\caption{Overseas traffic destinations of devices grouped by lifecycle phases. Device lifecycle phases: SU (setup), ID (idle), IR (interaction), DE (deletion). The destinations with values less than 0.1 are magnified. The vertical coordinate value is the ratio of the number of bytes flowing from the device to a certain country to the total number of bytes of the device's flow, multiplied by the number of devices.}
\label{Fig:lifecycle_destination}
\end{figure}

To analyze the distribution of traffic across the device lifecycle phases, we visualized the proportional traffic destination during the different lifecycle phases in Figure~\ref{Fig:lifecycle_destination}. Since over 99\% of the traffic terminates within China, the domestic traffic is omitted from the figure. Therefore, the traffic of plug, light, hub, humidifier, and sensor devices is also not included in the figure.
The results show that during the idle phase, devices exhibit a relatively high proportion of overseas traffic, and the traffic primarily flows to NTP services. 
Apart from the doorbell device, during the setup phase, all the devices show a significant amount of overseas traffic and involve diverse services, such as DNS service, IP address lookup, IoT platform management, and NTP service. This means the device configurations rely on many types of overseas services. 
During both the interaction and deletion phases, the frequency of overseas destinations decreases. This likely suggests that the main functionalities of IoT devices are built on domestic services.

Additionally, for 10 devices that updated firmware, none of them generated overseas traffic before or after the firmware updates.

\subsection{Servers contacted by devices}
\label{sec:4.2}

During device operation, frequent traffic communication occurs between devices and cloud service providers. We calculated the number of cloud servers belonging to different parties contacted by devices in different lifecycle phases. Table~\ref{table:Server Number} shows the average number of service providers contacted by devices of each type.

\begin{table}[h]
\huge
\centering
\caption{The average number of servers contacted by devices in different lifecycle phases.}
\label{table:Server Number}
\resizebox{0.9\textwidth}{!}{
\begin{tabular}{c|c|ccccccccc}
\toprule[4pt]
\textbf{Lifecycle}  & \textbf{Server} & \textbf{Camera} & \textbf{Doorbell} & \textbf{Hub} & \textbf{Humidifier} & \textbf{Light} & \textbf{Plug} & \textbf{Sensor} & \textbf{Speaker} & \multicolumn{1}{l}{\textbf{Other devices}} \\ \hline
                     & First party     & 3.24            & 2.33              & 0.56         & 2.00                & 1.33           & 0.83          & 1.33            & 9.22             & 13.50                                      \\
\textbf{Setup}       & Support party   & 2.10            & 1.00              & 0.78         & 0.00                & 0.33           & 0.67          & 0.00            & 0.89             & 0.00                                       \\
                     & Third party     & 1.34            & 0.67              & 0.00         & 0.00                & 0.00           & 0.00          & 0.00            & 0.67             & 6.00                                       \\ \hline
                     & First party     & 1.76            & 1.67              & 0.44         & 1.00                & 1.33           & 1.08          & 2.33            & 9.22             & 17.75                                      \\
\textbf{Idle}        & Support party   & 2.48            & 1.00              & 0.67         & 0.00                & 0.33           & 0.50          & 0.00            & 1.22             & 1.00                                       \\
                     & Third party     & 1.24            & 1.33              & 0.00         & 0.00                & 0.00           & 0.00          & 0.33            & 1.22             & 0.25                                       \\ \hline
                     & First party     & 1.38            & 4.00              & 0.33         & 1.00                & 0.67           & 0.58          & 1.33            & 3.89             & 11.50                                      \\
\textbf{Interaction} & Support party   & 1.69            & 1.33              & 0.22         & 0.00                & 0.17           & 0.25          & 0.00            & 1.33             & 0.75                                       \\
                     & Third party     & 0.90            & 1.67              & 0.00         & 0.00                & 0.00           & 0.00          & 0.33            & 1.00             & 5.25                                       \\ \hline
                     & First party     & 0.41            & 2.33              & 0.33         & 1.00                & 0.67           & 0.58          & 1.00            & 3.00             & 8.50                                       \\
\textbf{Deletion}    & Support party   & 0.59            & 0.00              & 0.33         & 0.00                & 0.17           & 0.17          & 0.00            & 0.22             & 0.75                                       \\
                     & Third party     & 0.21            & 1.00              & 0.00         & 0.00                & 0.00           & 0.00          & 0.33            & 0.22             & 0.00                                       \\ \hline
                     & First party     & 4.03            & 4.33              & 1.11         & 2.00                & 1.83           & 1.75          & 3.67            & 23.44            & 35.25                                      \\
\textbf{Total}       & Support Party   & 5.03            & 2.33              & 1.00         & 0.00                & 0.50           & 0.83          & 0.00            & 3.00             & 1.25                                       \\
                     & Third papty     & 4.00            & 2.00              & 0.00         & 0.00                & 0.00           & 0.00          & 0.33            & 2.33             & 7.00                                       \\ \toprule[4pt]
\end{tabular}}
\end{table}

First, regardless of the type of device, first party contacts constituted a significant proportion of all device communications. On average, each device contacts 7.0 first party servers, 2.6 support party servers, and 2.2 third party servers. This indicates that the privacy policies of Chinese devices are relatively well-established. Second, the interaction with the support party is closely related to the device's computational resource needs. Among the 12 device types, the camera, doorbell, speaker, and mirror interact more frequently with the support party. These devices, on average, contact 4.3 support party servers, while the remaining devices contact an average of 0.9 support party servers. Many manufacturers choose to collaborate with cloud service providers to ensure the devices can leverage cloud computing and storage capabilities, thereby enhancing their stability. The support party servers contacted by these devices typically provide cloud computing and storage services (e.g., oss-cn-beijing.aliyuncs.com, ap-beijing.myqcloud.com), video streaming services (e.g., video.qq.com, stream.qqmusic.qq.com), and network communication services (e.g., dns.alidns.com). 

Compared with first and support party servers, third party servers primarily involve data sharing, advertising services, geolocation, and personalized recommendation services. For example, the Huawei speaker contacts statist.tingmall.com for music streaming services, which may be related to third party audio components used by the device; the IMOU camera contacts api.weathercn.com for weather query services; the Huawei TV contacts several audio and video service providers, however, we did not find these service providers in their privacy policies. Additionally, we also observe that many devices access seemingly unrelated third party servers. For instance, Xiaomi speaker, Konka camera, Lenovo camera, Pisen camera, and Jooao camera all access the domain www.baidu.com. These observations suggest that while third party servers offer various personalized features, they are not explicitly mentioned in privacy policies. As a result, unexpected destination contacts may raise user privacy concerns.
Some third party servers possess strong cross-platform data aggregation capabilities. They can integrate the user behaviors on different devices to build detailed user behavior profiles, which brings serious privacy risks~\cite{iseeyoursmarthomeactivities,IoTMosaic}. Therefore, device manufacturers should clearly define data usage boundaries when integrating third party services, transparently disclose relevant third party domains and their purposes in privacy policies, and allow users to make choices.

Additionally, for 10 devices that support firmware updates, there was no significant change in the number of third party service providers contacted. Overall, traffic destinations showed little change after the firmware updates, indicating that device firmware updates have little impact on traffic destinations in the short term.

\textbf{Commonly contacted organizations.} To further assess whether the devices' communications exhibit excessive reliance on specific organizations, we analyzed the organization of traffic destinations and ranked them based on interaction frequency. We focused on the companies most frequently contacted by consumer IoT devices and listed the top ten companies in Table~\ref{table:Organization}. According to the table, 20 devices in the dataset contacted Google. By analysing the traffic, we find that devices contact Google to use its DNS service. In contrast, the Chinese DNS provider Greatbit, is used by only 17 devices, which is fewer than Google. 

In consumer IoT device real-time communication scenarios, stable and efficient DNS services are crucial. As a domestic service provider, GreatBit offers lower latency and higher stability for local devices, making it more suitable to meet the demands of domestic devices. Nevertheless, the analysis shows that compared with the local DNS service, the frequency of the device using Google DNS is higher. In addition to Google and Greatbit, the Organizations most frequently contacted by devices include Alibaba Cloud, Xiaomi, and Tencent. The services provided by these companies through their domains include cloud storage services (e.g., oss-cn-shenzhen.aliyuncs.com), NTP services (e.g., time1.cloud.tencent.com, ntp.aliyun.com), and video streaming and media content (e.g., findermp.video.qq.com, vv.video.qq.com), indicating that devices rely on these companies for the computational resources they provide.

\begin{table}[h]
\centering
\caption{Organizations commonly contacted by devices.}
\label{table:Organization}
\resizebox{\textwidth}{!}{
\begin{tabular}{c|cccccccccc}
\toprule[1.5pt]
\textbf{Organization} & Google & Alibaba Cloud & Xiaomi & Greatbit & Baidu & Tencent & Huawei & Tuya & NST & Jooan \\ \hline
\textbf{Frequency}    & 21     & 20            & 20     & 17       & 13    & 12      & 10     & 5    & 4   & 3 \\ \toprule[1.5pt] 
\end{tabular}}
\end{table}

\subsection{Summary}

By analyzing traffic destinations, we find that most Chinese devices rely on domestic services. Specifically, 99.75\% of the traffic is directed to domestic servers that depend on domestic organizations. This means that Chinese IoT devices feature a high degree of localization and greater autonomy in control. From the perspective of lifecycle phases, we can infer that the main functionalities of IoT devices, e.g., controlling and monitoring, are built on domestic services. Most overseas services used by Chinese consumer IoT devices are basic Internet services like NTP and DNS, which can also be replaced by Chinese service providers. 
In addition to contacting first parties, many devices from different manufacturers commonly communicate with major domestic organizations such as Xiaomi, Alibaba, and Tencent. This allows them to receive information from many devices and potentially obtain the capability to construct user profiles, e.g., inferring types of devices in a household. 

\section{Encryption analysis}
\label{sec:Encryption Analysis}

To ensure the security of consumer IoT device traffic, devices often encrypt the communication. However, encryption implementations vary across brands and device types, with differing security levels. In this section, we measure encryption practices through 4 key aspects: (1)the proportion of encrypted traffic, (2) the encryption protocols, (3) the certificate security in TLS, and (4) the exposure of personally identifiable information (PII) in unencrypted traffic.

\subsection{Encrypted traffic proportion analysis}

Consumer IoT device communications may involve sensitive information such as user personal data and surveillance video. Transmitting such data in plaintext raises interception, tampering, and unauthorized risks. Thus, we leverage the entropy of the traffic to identify and analyze the proportion of unencrypted~communications.  

Common traffic analysis tools, such as Wireshark, can easily distinguish encrypted protocols like TLS and QUIC, but cannot clearly classify customized encryption protocols. Therefore, we adapted the method proposed by Ren et al.~\cite{10.1145/3355369.3355577} to determine whether traffic is encrypted by calculating the payload entropy of the packets. First, for HTTP protocol packets, we classify them into text, compressed, or media based on the \emph{context\_type} field. For the SSL protocol, if the payload entropy is greater than 0.8, it is classified as encrypted. For the DNS protocol, packets containing the \emph{dns\_key} field are treated as encrypted. For packets that cannot be classified in the first step, we further classify them based on the \emph{magic\_number} of the payload. The \emph{magic\_number} refers to a unique identifier or pattern of bytes in a packet that helps to determine its type or format, often used to identify specific file types or data structures. If the packet is identified as compressed or media data, it is categorized as compressed or media. Finally, for the remaining packets, we perform a final classification based on entropy values: packets with entropy greater than 0.9 are classified as encrypted, those with entropy less than 0.4 are classified as text, and those with entropy between 0.4 and 0.9 are classified as unknown. Ultimately, we classify the traffic into encrypted traffic, unencrypted traffic, and unknown traffic.

\begin{figure}[h]
\centering
\includegraphics[width=1\linewidth]{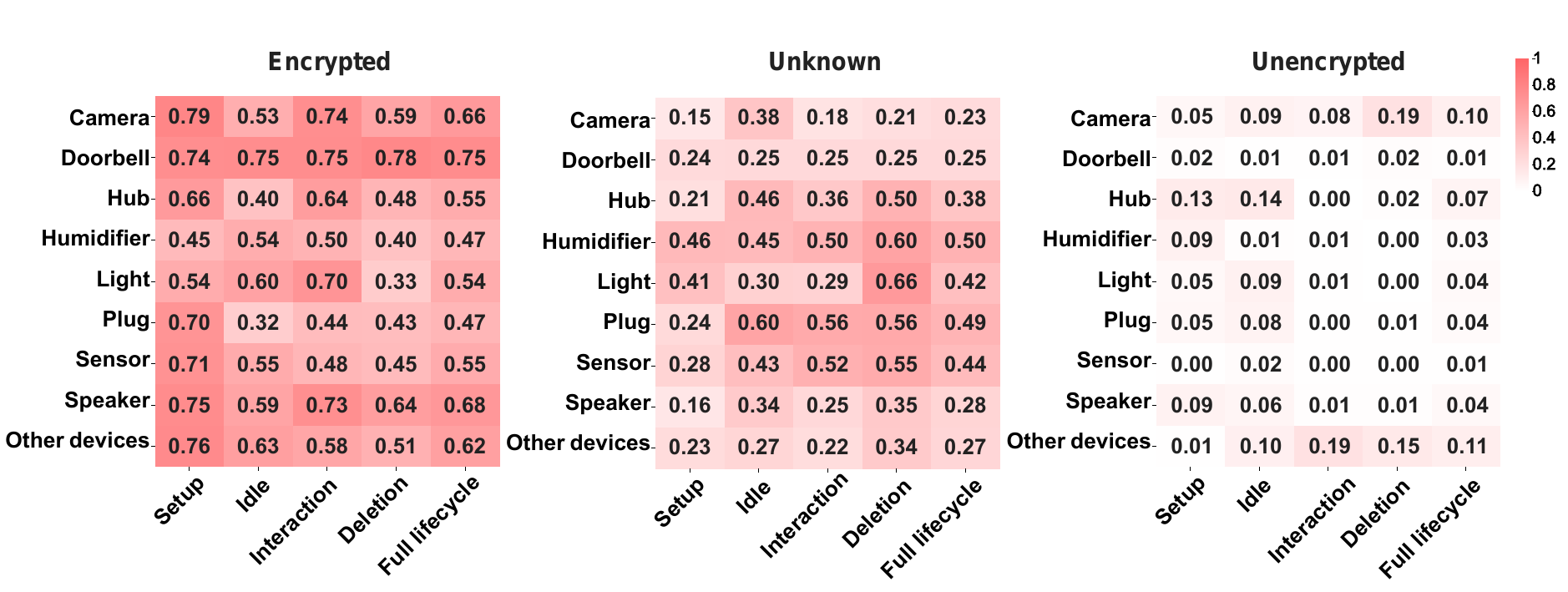}
\caption{For each device type, the heatmaps depict the average percentages of encrypted, unknown, and unencrypted traffic transmitted across respective lifecycle phases.}
\label{Fig:heat}
\end{figure}

Figure~\ref{Fig:heat} presents a heat map illustrating the proportional distribution of encrypted, unknown, and unencrypted traffic in various lifecycle phases of consumer IoT devices.

The heatmaps show that the proportion of unencrypted traffic is generally low, with no device type exceeding 11\% of unencrypted traffic in the full lifecycle. This indicates that encrypted communication dominates the data transmission of most devices. 
However, significant variation exists between different device types. For instance, camera, doorbell, speaker, and TV devices exhibit an average encrypted traffic share of over 60\%, largely driven by the security and privacy requirements in audio and video data transmission. 
In contrast, the humidifier, pet feeder, and plug devices have lower average encrypted traffic proportions of 36\%, 34\%, and 38\%, while their unknown traffic proportions are the highest among all devices, at 50\%, 54\%, and 49\%, respectively. Analysis of traffic entropy reveals that most of these devices’ traffic falls within the entropy range of 0.4-0.7. Upon further inspection, it appears that these devices do not use standard encryption protocols but instead rely on proprietary protocols developed by the manufacturers. These proprietary protocols have unstable entropy. 
By reverse engineering, we discovered that some devices can deduce the syntax and semantics of the manufacturer's protocols.
Therefore, we suggest that manufacturers adopt standardized encryption protocols, such as TLS 1.3. In cases where proprietary protocols must be used, they can be combined with packet padding and traffic obfuscation techniques~\cite{RandomSegmentation} to improve resistance against protocol reverse engineering and effectively reduce the risks of data leakage.

When considering the lifecycle phases of the device, the proportion of encrypted traffic is relatively higher during the setup and interaction phases than in the other two phases. These phases involve crucial services such as device initialization and configuration, cloud computing, and media data transmission. Therefore, more encrypted traffic in these phases ensures communication security.

Additionally, for 10 devices that support firmware updates, Most devices exhibited increased encrypted traffic proportions after updates. For example, TP-Link Camera increased the average encrypted traffic proportion from 65.75\% pre-update to 72.56\% post-update. This indicates that the encryption communication is becoming increasingly prevalent. 

\subsection{Encryption protocol analysis}

In addition to standard encryption protocols such as TLS and SSL, some manufacturers use proprietary encryption protocols based on UDP~\cite{paracha2021iotls}. Therefore, we first check whether standard encryption protocols, such as SSL or TLS, are used for encrypted traffic. Encrypted traffic not using these protocols is classified as encrypted with proprietary protocols. In the analysis, we statistic the number of devices using each encryption protocol with the results presented in Table~\ref{table:protocol}.

\begin{table}[h]
\centering
\caption{Distribution of encryption protocols deployed in devices.}
\label{table:protocol}
\resizebox{0.8\textwidth}{!}{
\begin{tabular}{c|ccccccc}
\toprule[1.5pt]
\textbf{Protocol}       & \textbf{TLS 1.1} & \textbf{TLS 1.2} & \textbf{TLS 1.3} & \textbf{SSL} & \textbf{SSLv2} & \textbf{SSLv3} & \textbf{Other protocols} \\ \hline
\textbf{Setup}          & 1                & 58               & 1                & 5            & 3              & 8              & 11                       \\
\textbf{Idle}           & 2                & 59               & 1                & 2            & 0              & 1              & 11                       \\
\textbf{Interaction}    & 4                & 54               & 3                & 18           & 14             & 12             & 11                       \\
\textbf{Delete}         & 0                & 48               & 0                & 1            & 0              & 2              & 11                       \\
\textbf{Full lifecycle} & 5                & 62               & 5                & 22           & 16             & 17             & 11                                             \\ \toprule[1.5pt]
\end{tabular}}
\end{table}

Through analysis, we found that TLS 1.2 is widely used for encrypted communication of devices, with 62 devices using TLS 1.2 at least once. 
However, some devices (e.g., IMOU doorbell, Pisen camera, and Tenda camera) continue to rely on insecure SSL or TLS 1.1 protocols, which account for 7.6\% of the total communication traffic on average. The devices generally use these protocols during the setup phase and the interaction phase. 
The setup and interaction phases often involve the transmission of sensitive data such as user identity information, remote control commands, and status update information. If weak encryption protocols are used during these phases, attackers may exploit them through MITM attacks to intercept and reconstruct the transmitted data.
Currently, protocols below TLS 1.2 have been deemed insecure by multiple international standards organizations. Devices that continue to use such protocols may fail to comply with data protection requirements such as the General Data Protection Regulation (GDPR) and China’s Cybersecurity Law. Therefore, device manufacturers should review and phase out insecure protocols such as SSL, TLS 1.0, and TLS 1.1, and adopt TLS 1.2 or higher for secure~communication.

Analyzing the destination IP addresses of the traffic showed that most targets are telecommunications service providers (e.g., China Mobile) or cloud service providers (e.g., Alibaba Cloud). 
Moreover, not all encrypted communications from these devices only use SSL or TLS 1.1. There is a mixed use with other protocols (e.g., TLS 1.2). This may result from manufacturers neglecting to fully deprecate outdated protocols during certain functionality upgrades, leaving some connections still using SSL or TLS 1.1. 
On the other hand, while TLS 1.3 offers significant improvements in performance and security, only five devices in our dataset use it. This indicates that despite its clear advantages, its adoption remains limited, and manufacturers must make greater efforts to implement the latest security protocols.
Additionally, we identified 11 devices (e.g., Aqara Camera, Aqara Plug, Dogness Feeder) using proprietary protocols developed by the manufacturers. The security of these protocols remains to be further analyzed.

In conclusion, although some devices still use outdated encryption protocols and custom encryption formats present security risks, TLS 1.2 remains the most widely used encryption protocol, and the adoption of TLS 1.3 requires more attention from manufacturers in future updates.

Additionally, for the 10 devices that support firmware updates, most used TLS 1.2 for encryption before the update, while three devices relied on proprietary protocols developed by the manufacturers. The XiaobaiY2 camera also used SSL in addition to TLS 1.2.
After the update, all devices upgraded their encryption protocols. In addition to continuing to use TLS 1.2, they also adopted TLS 1.3. The three devices that previously used proprietary protocols switched to use TLS 1.2, and the XiaobaiY2 camera stopped using SSL, now relying on TLS 1.2 and TLS 1.3 for encrypted communication.

\subsection{TLS certificate security analysis} 

The TLS 1.2 protocol is the most widely used encryption protocol in our dataset. 
To further understand its security, this subsection presents a statistical analysis of the certificate for the 62 devices in the dataset that use the TLS 1.2 protocol. 
The results show that 48 devices use a public key certificate model for communication. The remaining 14 devices use the Pre-Shared Key (PSK) model without relying on certificates.
Given the crucial role digital certificates play in TLS encryption security, we further analyzed the self-signed certificate usage and the security of the encryption algorithms. Additionally, we conducted dynamic tests by using mitmproxy to assess the validity of certificates when devices are powered on again.

\subsubsection{Certificate usage} 
Overall, the certificate security of devices is generally good, but some insecure certificates were still identified. Table~\ref{table:certificate} statistics the number of TLS implementations that may be insecure, including insecure signature algorithms, insecure cryptographic algorithms, and self-signed certificates. 
Among the 48 devices, 2 use insecure algorithms, such as the SHA1 signature algorithm and the RSA-1024 cryptographic algorithm, primarily for communication with Aliyun and AWS servers. These algorithms are now considered insecure and vulnerable to attacks. 
Therefore, device manufacturers can regularly audit the cypher suites and certificate configurations used in their products, eliminate algorithms that are widely recognized as insecure, and ensure that their TLS implementations comply with current industry security~standards.

Secondly, approximately 37.5\% of the devices use self-signed certificates in at least one TLS 1.2 connection. These certificates are typically signed by CA authorities like ClickPKI or self-signed by the device manufacturers. Many self-signed certificates have excessively long validity periods, with one Lenovo camera certificate valid until 2120. According to current security standards, the validity period of TLS certificates should not exceed 13 months~\cite{TLScertificate}. As pointed out by Dong et al.~\cite{dong2023behind}, the lack of public oversight for long-term vendor-signed certificates poses potential vulnerabilities, as they may not meet the stringent standards of trusted public CAs, thereby increasing the security risks within the consumer IoT ecosystem.

\begin{table}[h]
\centering
\large
\caption{TLS 1.2 implementations of different device type. }
\label{table:certificate}
\resizebox{\textwidth}{!}{
\begin{tabular}{c|cccccccccccc}
\toprule[1.5pt]
\multirow{2}{*}{\textbf{TLS   1.2 implementation}} & \multicolumn{3}{c|}{\textbf{Camera}}                         & \multicolumn{3}{c|}{\textbf{Doorbell}}                       & \multicolumn{1}{c|}{\textbf{Plug}} & \multicolumn{1}{c|}{\textbf{Hub}} & \multicolumn{1}{c|}{\textbf{Sensor}} & \multicolumn{1}{c|}{\textbf{Speaker}} & \multicolumn{2}{c}{\textbf{Other devices}} \\ \cline{2-13} 
                                                   & \textbf{SU} & \textbf{ID} & \multicolumn{1}{c|}{\textbf{IR}} & \textbf{ID} & \textbf{IR} & \multicolumn{1}{c|}{\textbf{DE}} & \multicolumn{1}{c|}{\textbf{ID}}   & \multicolumn{1}{c|}{\textbf{ID}}  & \multicolumn{1}{c|}{\textbf{ID}}     & \multicolumn{1}{c|}{\textbf{SU}}      & \textbf{SU}          & \textbf{DE}         \\ \hline
\textbf{Insecure   signature algorithms}           & 1           & 0           & 0                                & 0           & 0           & 0                                & 0                                  & 1                                 & 0                                    & 0                                     & 0                    & 0                   \\
\textbf{Insecure   cryptographic algorithms}       & 1           & 0           & 0                                & 0           & 0           & 0                                & 0                                  & 1                                 & 0                                    & 0                                     & 0                    & 0                   \\
\textbf{Self-signed   certificates}                & 10          & 3           & 2                                & 1           & 1           & 1                                & 1                                  & 1                                 & 1                                    & 1                                     & 2                    & 1                   \\ \toprule[1.5pt]
\end{tabular}}
\begin{tablenotes}   
        \footnotesize               
        \item[4] *Device lifecycle phases: SU (setup), ID (idle), IR (interaction), DE (deletion).
\end{tablenotes} 
\end{table}

\vspace{-1em}
\subsubsection{Certificate validation} 

Ideally, devices should check whether the certificates from the servers are valid. If a forged certificate is detected, the device should either return an error message or reject the TLS connection. However, if a device ignores certificate anomalies and proceeds to establish a connection, the certificate validation mechanism may have vulnerabilities.

To investigate this issue, we first captured the traffic during device reconnections. We discovered that 40 devices conduct TLS connections with certificate exchange with 106 servers during the reconnection period. Next, we conducted a mitmproxy certificate replacement attack on these TLS connections. The results are presented in Table~\ref{tab:mitm_results}.

\begin{table}[htbp]
  \centering
  \caption{Error messages or device behavior after the certificate replacement}
  \label{tab:mitm_results}
  \resizebox{0.7\textwidth}{!}{
  \begin{tabular}{c|cc}
    \toprule[1.5pt]
    \textbf{Error message and device behavior} & \textbf{Device number} & \textbf{Server number} \\
    \hline
    ``Unknown CA'' & 14 & 25 \\
    ``Decrypt error'' & 2 & 2 \\
    ``Bad certificate'' & 2 & 2 \\
    ``Close notify'' & 5 & 15 \\
    ``Decode error'' & 1 & 1 \\
    Disconnects and reconnects & 12 & 19 \\
    Server handshake failed & 2 & 3 \\
    Unable to connect to the internet & 1 & 1 \\
    Communicates normally & 23 & 38 \\
    \toprule[2pt]
  \end{tabular}}
\end{table}

The results show that approximately 65\% of the servers responded with errors, such as "Unknown CA," "Decrypt error," "Close notify" alerts, or actively terminated the connection, and 35\% of the servers associated with 23 devices communicate normally. 
We decrypted the traffic between the devices and servers and extracted APIs used by the devices, revealing critical device information. For example, the API ``/S\_PROD1\_MEIDIMINI/.../215413.log'' exposes device log data, the API ``/v3/am/query\_device\_bind\_\\status'' reveals device binding status, and in this API, the fields ``device\_name'' and ``device\_sk'' are exposed. We suspect that ``device\_sk'' may be related to the device's secret key, which is typically used to verify the identity and ensure the security of the device. These APIs expose sensitive information, if exploited by attackers, they could lead to serious consequences such as log data leakage, unauthorized remote control of the device, or tampering with its binding status. These findings suggest that many consumer IoT devices adopt insecure certificate validation mechanisms, making them vulnerable to certificate replacement attacks. We have reported this issue to the respective device manufacturers and submitted the vulnerability to the China National Vulnerability Database (CNVD).

\subsection{PII exposure in unencrypted traffic}

Personally Identifiable Information (PII) refers to any data that can identify an individual~\cite{PIIDefinitions}. Common PII includes names, ID numbers, phone numbers, email addresses, etc. Unencrypted traffic containing PII significantly elevates the risk of user data exposure. In this subsection, we identify PII in unencrypted traffic using text matching and regular expressions to measure the security practice. To comprehensively analyze PII exposure, we input as much personal information as possible into the companion app when we collect traffic. 

The analysis reveals that the exposure of PII in unencrypted communication traffic is minimal. Only a small amount of PII (including the names of five well-known singers) is detected in the traffic from four speaker devices, and no inputted personal information is found. These singer names appeared in the song requests during the interaction phase. Although this information does not directly reveal the user's identity, it could still be exploited by malicious third parties to infer user preferences or deliver targeted ads, thereby posing a risk of privacy leakage and data misuse.

\subsection{Summary}

This section measures the encryption practices of consumer IoT devices. The study reveals that Chinese manufacturers adopt safe encryption protocols for device communications, and encrypted traffic dominates most communications. However, some devices still use deprecated protocols, though their traffic constitutes a small proportion(7.6\%), and some devices use proprietary protocols developed by the manufacturers rather than standardized protocols, bringing about certain security risks. Encouragingly, following firmware updates, 10 devices adopted more secure encryption protocols after the update, indicating a positive trend toward encryption protocol usage. Most devices use secure certificates, but prevalent self-signed certificates and insecure algorithms persist as potential risks. In terms of certificate verification, many devices have the problem of insecure certificate verification mechanisms, which may also bring potential risks. While direct PII exposure in unencrypted traffic is rare, inferred user preferences could enable targeted advertising. These findings indicate that the devices should adopt standard encryption protocols and reduce reliance on self-signed certificates to further minimize risks.

\vspace{-1em}

\section{Comparisons on different regions}
\label{sec:Comparisons on Different Regions}

This section compares consumer IoT device traffic measurement results in China with those in other regions.  We choose Ren et al.’s~\cite{10.1145/3355369.3355577} and Paracha M. T. et al.’s~\cite{paracha2021iotls} for comparison, as their studies align with our research purpose, focusing on traffic destinations and encryption analysis.

\subsection{Traffic destinations}

We have summarized the differences of traffic destinations between China and other regions in the table~\ref{table:compare_des}. In terms of traffic destinations, Chinese devices mainly communicate domestically when compared with consumer IoT devices in the United States and the United Kingdom.

\begin{table}[h]
\captionsetup{labelformat=empty}
\caption{Table 7: Comparison of traffic destinations with other regions.}
\label{table:compare_des}
\centering
\begin{threeparttable}
\resizebox{\textwidth}{!}{
\begin{tabular}{cccc|ccc}
\toprule[2pt]
\multicolumn{4}{c|}{\textbf{Distribution of traffic destinations}\tnote{1}}                                                                          & \multicolumn{3}{c}{\textbf{Commonly contacted   organizations}}                  \\ \hline
\multicolumn{1}{c|}{\textbf{destination}}               & \textbf{China (77)} & \textbf{United States (46)} & \textbf{United Kingdom (35)} & \textbf{China (77)} & \textbf{United States (46)} & \textbf{United Kingdom (35)} \\ \hline
\multicolumn{1}{c|}{\textbf{China}}          & 97.00\%            & 2.40\%                     & 1.69\%                      & Google             & Amazon                     & Amazon                      \\
\multicolumn{1}{c|}{\textbf{United States}}  & 2.70\%             & 90.00\%                    & 72.90\%                     & Alibaba Cloud      & Google                     & Google                      \\
\multicolumn{1}{c|}{\textbf{Europe}}         & 0.20\%             & 7.50\%                     & 22.03\%                     & Xiaomi             & Akamai                     & Akamai                      \\
\multicolumn{1}{c|}{\textbf{Other Regions}} & 0.10\%             & 0.10\%                     & 3.38\%                      & Greatbit           & Microsoft                  & Microsoft                   \\ \cline{1-4}
\multicolumn{4}{c|}{\textbf{Service Contacted by Devices}}                                                                               & Baidu              & Netflix                    & Alibaba                     \\ \cline{1-4}
\multicolumn{1}{c|}{\textbf{party}}               & \textbf{China (77)} & \textbf{United States (46)} & \textbf{United Kingdom(35)} & Tencent            & Kingsoft                   & Kingsoft                    \\ \cline{1-4}
\multicolumn{1}{c|}{\textbf{Support party}}  & 2.68               & 2.13                       & 2.48                        & Huawei             & 21Vianet                   & 21Vianet                    \\
\multicolumn{1}{c|}{\textbf{Third party}}    & 1.71               & 0.15                       & 0.14                        & Tuya               & Alibaba                    & Beijing Huaxiay              \\ \toprule[2pt]
\end{tabular}}
\begin{tablenotes}    
        \footnotesize               
         \item[1] The ratio of the United States and the United Kingdom are estimated in Ren et al.'s~\cite{10.1145/3355369.3355577} work.
\end{tablenotes} 
\end{threeparttable} 
\end{table}

According to Ren et al.~\cite{10.1145/3355369.3355577}, the 35 devices in the United Kingdom communicated with 7 overseas regions. Approximately 70\% of the United Kingdom's device traffic flowed outside these regions (primarily to the United States). 
In contrast, although Chinese consumer IoT devices connected with 18 overseas countries, overseas traffic only accounted for 3\% of total bytes. This highlights Chinese consumer IoT devices’ stronger reliance on domestic services than their United Kingdom counterparts. 
The devices in the United Kingdom partially rely on overseas cloud platforms and service providers (such as AWS, Google Cloud, and Akamai), resulting in frequent cross-border transmission of their data to overseas data centres. This increases the risks of data leakage. 
In contrast, most data transmission activities of Chinese consumer IoT devices occur within China, indicating a high degree of domestic service deployment. This result reflects the high degree of reliance of Chinese manufacturers on domestic services.

Regarding commonly contacted organizations, Chinese consumer IoT devices mainly interact with domestic providers. 
In Ren et al.'s study~\cite{10.1145/3355369.3355577}, the United Kingdom devices are mainly connected to the United States organizations, such as Amazon, Google, and Akamai. 
In our analysis, except for Google, Chinese devices primarily communicated with domestic organizations such as Alibaba Cloud, Xiaomi, and Tencent. These providers offer diverse services, including IoT platform services, cloud storage services, and video streaming and media content (See Section~\ref{sec:4.2} for details). 

At last, according to the findings of Ren et al.~\cite{10.1145/3355369.3355577}, the number of third parties contacted by devices is generally low, with the setup phase involving more third parties than other phases, likely due to devices establishing initial connections during this phase. 
In our study, the number of third parties contacted by devices is also low, suggesting that devices in different regions tend to communicate more with first and support parties. 
On the other hand, we find that the number of third parties contacted during the idle phase is higher than in other phases, which differs from Ren et al.'s findings. Devices in the United States and the United Kingdom tend to complete third party dependency initialization during the setup phase, while Chinese devices are more likely to maintain long-term connections with third parties during the idle phase for purposes such as data uploading, status synchronization, or push services.

\subsection{Encryption analysis}

Encryption practices vary regionally, but unencrypted traffic proportions remain low overall. 
Ren et al.~\cite {10.1145/3355369.3355577} proposed entropy-based encryption detection, reporting unencrypted traffic proportions below 13\% for devices in the United States and the United Kingdom. 
Using similar methods, we come to a similar conclusion: the percentage of unencrypted traffic of consumer IoT devices in China is low, with a peak of 11\%. 

Regarding encryption protocol usage, TLS 1.2 remains the most widely adopted, while TLS 1.3 adoption is still limited. 
Paracha M. T. et al.~\cite{paracha2021iotls} analyzed 40 consumer IoT devices in the United States in 2021 and found that TLS 1.2 is the dominant protocol, with minimal use of TLS 1.3 and deprecated versions. 
Our findings are similar to these results. Among 77 Chinese devices, 81\% utilized TLS 1.2, while 7.6\% still relied on insecure protocols (TLS 1.1/SSL), and 14\% employed proprietary protocols developed by manufacturers.
For certificate validation, Paracha M. T. et al.~\cite{paracha2021iotls} examined 32 devices and found that 11 were vulnerable to certificate replacement. 
In our study, 17 out of 40 devices generated error messages when the certificate was replaced, while 23 devices exhibited no errors or anomalies, indicating that certificate verification remains a security concern.

In terms of PII exposure, Ren et al.'s study~\cite{10.1145/3355369.3355577} identified PII in unencrypted traffic, including MAC addresses, UUIDs, device IDs, geolocation data, and user-assigned device names. 
In contrast, Chinese devices exposed minimal PII, only singer names in the speaker's traffic, indirectly revealing music preferences. This indicates stronger privacy protection in Chinese devices’ unencrypted traffic.

\subsection{Summary}

A clear regional difference is observed in the traffic destinations and encryption practices. Chinese consumer IoT devices are more reliant on domestic service providers, in contrast to devices in the United Kingdom, which mainly rely on international cloud platforms such as Amazon. This suggests that China has developed a more self-reliant and locally integrated IoT ecosystem, with limited dependence on international platforms.
Encryption practices and certificate implementations are generally consistent across regions, ensuring relative security, but a certain proportion of devices still use deprecated protocols, which could pose potential security and privacy risks. Meanwhile, there are weaknesses with certificate verification in devices from different regions, posing a security concern. The PII contained in the unencrypted traffic of Chinese consumer IoT devices is less than that of their counterparts in the United Kingdom and the United States.
\vspace{-1em}

\section{Related work}
\label{sec:Related Work}

\textbf{The measurement of IoT traffic.}
Researchers have conducted a lot of measurements on IoT traffic, mainly focusing on security and privacy. Security measurements involve mapping IoT backend servers~\cite{saidi2022deep}, analyzing the impact of botnets on devices~\cite{noroozian2021can,herwig2019measurement}, and investigating data leakage and device vulnerabilities through companion apps~\cite{wang2019Looking,YuhongNan2023areyou,chen2018iotfuzzer,Redini2021Diane}. Said et al.~\cite{saidi2022deep} mapped IoT backend servers by analyzing ISP passive traffic data, revealing relationships between backend providers. Arman et al.~\cite{noroozian2021can} assessed the impact of two ISP security policies on the Mirai botnet. Stephen et al.~\cite{herwig2019measurement} studied the Hajime botnet’s effect on devices using active scanning and passive DNS backscatter traffic collection. These studies primarily focus on device-level evaluations while overlooking the user perspective. Privacy measurements include traffic destination identification~\cite{mandalari-pets21}, user data collection analysis~\cite{girish-imc23,dubois-pets20}, and the effectiveness of existing privacy protection policies~\cite{mandalari-sp23}. Mandalari et al.~\cite{mandalari-pets21} proposed an automated method to identify IoT traffic destinations, distinguishing between essential and non-essential ones. Aniketh et al.~\cite{girish-imc23} examined privacy leaks in local communication models using standard protocols. Dubois et al.~\cite{dubois-pets20} investigated how, when, and why smart speakers record external audio, while Anna Maria et al.~\cite{mandalari-sp23} tested the effectiveness of existing privacy protection measures. Compared with these studies, our work directly analyzes IoT device security practices by evaluating traffic destinations and encryption practices. Ren et al.~\cite{10.1145/3355369.3355577} and Paracha et al.~\cite{paracha2021iotls} have similar research purposes to ours. Paracha et al.~\cite{paracha2021iotls} evaluated the security of TLS connections in IoT devices, exploring how these connections evolve over time and whether certificates are properly validated. Ren et al.~\cite{10.1145/3355369.3355577} analysis the information exposure of devices in the United Kingdom and the United States. Inspired by their work, we introduce an analysis of Chinese devices and compare the results with other regions.

\begin{table}[h]
\caption{The summary of IoT device datasets.}
\label{table:dataset}
\centering
\begin{threeparttable}
\resizebox{0.9\textwidth}{!}{
\begin{tabular}{cccccccccccc}
\toprule[1.5pt] 

\multirow{2}{*}{Name} & \multirow{2}{*}{Area\tnote{1}} & \multirow{2}{*}{Source\tnote{2}} & \multirow{2}{*}{Categories} & \multirow{2}{*}{Number(IoT)\tnote{3}} & \multirow{2}{*}{Period} & \multirow{2}{*}{Size} & \multirow{2}{*}{Time} & \multicolumn{4}{c}{Lifecycle\tnote{4}} \\ \cline{9-12} 
                      &                       &                         &                             &                              &                         &                       &                       & SU    & ID    & IR    & DE    \\ \toprule
Ours~\cite{osti_10314043}                  & US                    & SC                      & 10                          & 8                            & 2020.3                  & 11.5GB                & 11 days               & \ding{55}     & \checkmark     & \checkmark     & \ding{55}     \\
YourThings~\cite{alrawi2019sok}            & US                    & SC                      & 15                          & 45                           & 2018.3                  & 233GB                 & 13 days               & \ding{55}     & \ding{55}     & \checkmark     & \ding{55}     \\
IoTDNS~\cite{9230403}                & US                    & SC                      & 28                          & 53                           & 2019.8                  & 366MB                 & 2 months              & \ding{55}     & \ding{55}     & \checkmark     & \ding{55}     \\
UNSW~\cite{8440758}                 & AUS                   & SC                      & 17                          & 28                           & 2016.10                 & 9.72GB                & 6 months              & \ding{55}     & \checkmark     & \checkmark     & \ding{55}     \\
BoT-IoT~\cite{koroniotis2019towards}               & AUS                   & SL                      & 5                           & 5                            & 2018.4                  & 69.3GB                & 2 months              & \ding{55}     & \checkmark     & \checkmark     & \ding{55}     \\
Mon(IoT)r~\cite{10.1145/3355369.3355577}             & US\&UK                & SC                      & 15                          & 81                           & 2018.9                  & 12.9GB                & -                     & \ding{55}     & \checkmark     & \checkmark     & \ding{55}     \\
PingPong~\cite{trimananda2020packet}              & US                    & SC                      & 12                          & 19                           & 2019                    & 40.3GB                & 51 days               & \ding{55}     & \ding{55}     & \checkmark     & \ding{55}     \\
HomeSnitch~\cite{campos2021towards}            & US                    & SC                      & 13                          & 57                           & 2021.3                  & 595MB                 & 8 days                & \ding{55}     & \ding{55}     & \checkmark     & \ding{55}     \\
IoT\_Sentinel~\cite{7980220}         & FI                    & SC                      & 6                           & 31                           & 2016                    & 61.4MB            &  -  & \checkmark                     & \ding{55}     & \ding{55}     & \ding{55}        \\
IoT23~\cite{bhandari2023distributed}                 & CZ                    & SC                      & 3                           & 3                            & 2018                    & 21GB                  & 1 year                & \ding{55}     & \ding{55}     & \checkmark     & \ding{55}     \\
N-BaIoT~\cite{meidan2018n}               & IL                    & SC                      & 3                           & 9                            & 2018.3                  & 240GB             &  -  & \ding{55}                     & \ding{55}     & \checkmark       &    \ding{55}   \\
NSL-KDD~\cite{tavallaee2009detailed}               & US                    & SL                      & -                     &-      & 1998.5                       & 4.06MB                  & 7 weeks            & \ding{55}                     & \checkmark     & \checkmark     & \ding{55}       \\
\textbf{CN-CIoT}            & CN                    & SC                      & 12                           & 77                           & 2024.3                  & 3GB                 & 4.5 months              & \checkmark     & \checkmark     & \checkmark     & \checkmark     \\

\toprule[1.5pt]
\end{tabular}}

\begin{tablenotes}    
        \footnotesize               
         \item[1] US (United States), UK (United Kingdom), AUS (Australia), FI (Finland), CZ (Czech Republic), IL (Israel), CN (China).
        \item[2] Data collection method labels: SC (Self collection), CR (Crowd sourcing), SL (Device simulation).
        \item[3] The number of IoT devices in the dataset.
        \item[4] Device lifecycle phases: SU (setup), ID (idle), IR (interaction), DE (deletion).
\end{tablenotes} 
\end{threeparttable} 
\end{table}

\textbf{IoT dataset construction.}
The IoT community has developed diverse datasets to facilitate research in consumer IoT devices. Table~\ref{table:dataset} summarizes the most highly cited datasets. Among self-collected datasets, Mon(IoT)r~\cite{10.1145/3355369.3355577} is the most cited and largest in device count, followed by UNSW ~\cite{8440758}, YT ~\cite{zavalyshyn2022sok}, Ours~\cite{osti_10314043}, and PingPong ~\cite{trimananda2020packet}. 
Previous studies have provided valuable datasets for consumer IoT device research, but there are still some limitations. First, these datasets were collected before 2021, lacking recent consumer IoT traffic. Second, most existing datasets focus on device traffic from Europe and the United States, while traffic from Asia is rare. Third, current datasets do not fully consider the entire lifecycle of devices during the collection, restricting the analysis from the perspective of the lifecycle. In contrast, our dataset includes traffic with the entire lifecycle of the device, enabling researchers to conduct more fine-grained analysis of consumer IoT device traffic.

In summary, we construct the first Chinese consumer IoT traffic dataset that covers the entire lifecycle of devices, completing the research gap in the dataset for China. This dataset enables a more fine-grained analysis of traffic destinations and encryption practices, and compares our measurements with those of other regions. By constructing the dataset and conducting measurements, this study provides new perspectives and insights for future research.

\vspace{-1em}

\section{Conclusion}
\label{sec:Conclusion}

This study constructs the largest dataset of consumer IoT device traffic in China, proposes a fine-grained traffic collection guidance covering the entire lifecycle of the device, and conducts measurements on traffic destination and encryption practice. 

We collected 108 hours of traffic based on 77 consumer IoT devices across 12 categories and 38 brands. Through analysis, we draw the following key conclusions. First, 99.75\% of traffic from Chinese consumer IoT devices is routed domestically, with only 0.25\% directed overseas. In contrast, approximately 70\% of the United Kingdom's device traffic flowed overseas. This indicates minimal reliance on overseas services. However, a notable observation is that most devices depend on services from major corporations like Alibaba Cloud, Xiaomi, and Tencent, indicating that most devices rely on the services of these organizations, and these organizations may also obtain more user information from devices. Regarding encryption practices, Chinese consumer IoT devices performed relatively well. The unencrypted traffic of Chinese devices peaked at 11\%, lower than the other regions, and Chinese devices have less PII exposure in unencrypted traffic. Nevertheless, some devices still employ deprecated encryption protocols, insecure TLS certificate algorithms (e.g., SHA1, RSA-1024), and self-signed certificates. More critically, during device certificate replacement, certain devices exhibited an absence of certificate validation mechanisms, exposing communication to potential security and privacy risks. 

We have open-sourced our traffic collection guidance and make our dataset publicly available on the website\footnote{\url{https://github.com/NKUHack4FGroup/Lifecycle-Based-Traffic-Dataset}} to facilitate future research.

\Acknowledgements{This work was partly supported by the National Natural Science Foundation of China No.62102198, the Fundamental Research Funds for the Central Universities No.079-63253221, and the Key Program of the National Natural Science Foundation of China under No.62032012 and No.62432012.
}










\end{document}